\documentclass[aps,prb,twocolumn,superscriptaddress]{revtex4-2}

\usepackage{amsmath,amssymb,graphicx,bm}

\usepackage{titlesec}
\usepackage{appendix}

\usepackage{xcolor}

\usepackage{hyperref}
\hypersetup{
	unicode=true,           
	colorlinks=true,       	
	linkcolor=blue,
	citecolor=blue,
	filecolor=magenta,
	urlcolor=blue
}

\usepackage{orcidlink}

\newcommand{\paperdoi}[1]{%
	\par\smallskip
	\noindent DOI: \href{https://doi.org/#1}{#1}%
	\par\medskip
}

\long\gdef\COMMENT#1{}
\begin{document}
	
	\title{Effect of charge-imbalance potential relaxation on the high-frequency vortex dynamics and kinetic inductance of superconducting circuits}
	\author{R.~I.~Kinzibaev\,\orcidlink{0009-0001-1404-7198}}
	\affiliation{Moscow Institute of Physics and Technology (MIPT), Dolgoprudnyi, Moscow region, 141700 Russia}
	\affiliation{Dukhov Research Institute of Automatics (VNIIA), 127055 Moscow, Russia}
	\author{V.~S.~Stolyarov\,\orcidlink{0000-0002-5317-0818}}
	\affiliation{Moscow Institute of Physics and Technology (MIPT), Dolgoprudnyi, Moscow region, 141700 Russia}
	\affiliation{Dukhov Research Institute of Automatics (VNIIA), 127055 Moscow, Russia}
	\author{A.~S.~Mel'nikov\,\orcidlink{0000-0002-4241-467X}}
	\affiliation{Moscow Institute of Physics and Technology (MIPT), Dolgoprudnyi, Moscow region, 141700 Russia}
	\affiliation{Institute for Physics of Microstructures, Russian Academy of Sciences, 603950 Nizhny Novgorod, GSP-105, Russia}
	
	\begin{abstract}
		We show that relaxation of the charge-imbalance potential plays a key role in the retarded dynamics of Abrikosov vortices in a type-II superconducting sample carrying a microwave current. Starting from the time-dependent Ginzburg--Landau equations we derive the vortex equation of motion accounting both the dissipation and retardation effects. The retardation is governed by the dynamics of the charge-imbalance potential and reveals itself at characteristic timescales diverging near the superconducting critical temperature $T_{c}$. These retardation effects in vortex dynamics strongly affect the kinetic inductance of superconducting circuits being, thus, responsible for the magnetic field dependence of characteristics of different superconducting devices in the high frequency range.
		\\
		\paperdoi{10.1103/l9n7-v4x5}
	\end{abstract}
	
	\maketitle
	
	\section{INTRODUCTION}\label{sec:i}
	
	Studying effective ways of tuning the kinetic inductance $L_{k}$ in superconducting circuits is an active area of research which can provide important recommendations for design of different cryoelectronic devices including superconducting resonators \cite{GopplM2008}, detectors of electromagnetic radiation \cite{ZmuidzinasJ2012} and qubits \cite{KrantzP2019}. For example, recent theoretical and experimental studies have demonstrated that magnetic control of superconducting transport properties and kinetic inductance in hybrid structures can be used to tune the parameters of superconducting electronic elements \cite{BakurskiyS2020,NeiloAA2025,NazhestkinIA2025}.
	
	The underlying tuning mechanisms can be based, e.g., on the suppression of the superfluid density $n_{s}$ by the applied magnetic field $\mathbf{B}$ or supercurrent $I_{s}$ which, in turn, lead to an increase in the magnetic field penetration depth $\lambda$:
	\begin{equation}\label{lk}
		L_k \propto \lambda^{2} \propto \frac{1}{n_s(\mathbf{B},I_{s})}.
	\end{equation}
	
	The related changes in the kinetic inductance and resistivity directly affect the resonance frequency and quality factor of superconducting microwave resonators and provide a basis for the operation of the kinetic-inductance detectors (KIDs) \cite{SemenovAD2001, VodolazovDYu2003, DayPK2003, RoitmanA2024No1, RoitmanA2024No2, MooreDC2009,VardulakisG2008}. Kinetic inductance tuning is also employed in nonlinear resonators coupled to superconducting qubits to facilitate readout of qubit states \cite{VissersMR2015, BoissonneaultM2012, KrantzP2016, PuriS2017, StockklauserA2017, LandigAJ2018}. Furthermore, sensitivity of $L_{k}$ to different perturbations can be used in high-sensitivity magnetometers \cite{LuomahaaraJ2014} and parametric amplifiers \cite{Castellanos-BeltranMA2007, HoEomB2012, SiddiqiI2004, MalnouM2021, ChienWC2023}.
	
	One can distinguish two basic mechanisms underlying the effect of applied magnetic field   on the kinetic inductance: (i) first, the field-induced supercurrents suppress the superconducting order parameter and, thus, cause the increase in the London penetration depth according to Eq.~\eqref{lk}; (ii) second, rather strong magnetic fields (depending on the sample geometry) can cause the penetration of Abrikosov vortices into the sample and the corresponding changes in the kinetic inductance in this regime will be also affected by the vortex dynamics. 
	
	Different models of vortex dynamics have been studied for more than half a century starting from the seminal works of Bardeen and Stephen  \cite{StephenMJ1965, BardeenJ1965}, where the vortex was treated as a viscous particle driven by the Lorentz force $\mathbf{F}_{L}$: 
	\begin{equation}\label{eq-int-viscous-motion}
		\eta \mathbf{v} = \mathbf{F}_{L},
	\end{equation}
	where $\mathbf{v}$ is the vortex velocity. The viscosity coefficient $\eta$ is determined by the dissipation processes in the vortex core region and can be evaluated for an isolated vortex line for low vortex densities when the intervortex distance well exceeds the core size (of the order of superconducting coherence length $\xi$). More detailed analyzes based on the time-dependent Ginzburg--Landau equations (TDGL) \cite{GorkovLP1975} or microscopic treatments \cite{KopninNB2001} have shown that the viscosity $\eta$ consists of two parts, $\eta_{\text{Rel}}$ and $\eta_{\text{Ohm}}$, which are associated with the relaxation of the superconducting order parameter and dissipative normal currents, respectively. However, the response of a superconductor in the mixed state to an external ac drive (e.g., a microwave transport current) cannot be captured by a purely viscous description of vortex motion. The {f}requency-dependent part of the vortex response can appear in the above equation of motion provided that we take account of the additional forces bounding the vortex to the pinning centers and the inertial effects described by the vortex mass (m) term:
	\begin{equation}\label{eq:of:motion:with:m:and:k}
		\left( -i \omega m + \eta - \eta \dfrac{\omega_{k}}{i\omega} \right) \tilde{\mathbf{v}} (\omega) = \tilde{\mathbf{F}}_{L},
	\end{equation}
	where $\tilde{\mathbf{v}} (\omega)$ is the vortex velocity in the frequency domain, $\tilde{\mathbf{F}}_{L} = ( \Phi_{0} / c ) [\tilde{\mathbf{j}}_{s,\infty} {\times} \mathbf{e}_{z}]$, $\omega_{k}$ is the depinning frequency (see, e.g., \cite{GolosovskyM1994}), the term $-\eta \omega_{k}/(i\omega)$ originates from the pinning force $- \eta \omega_{k} \mathbf{r}$ acting on a vortex in a parabolic potential and $\mathbf{r} = \tilde{\mathbf{v}} / (- i \omega)$ in Fourier representation.
	The nonstationary vortex dynamics and related microwave response in the presence of pinning were analyzed theoretically, e.g., \cite{CoffeyMW1991} by Coffey and Clem. However, their approach accounted only for the extrinsic mechanisms, such as pinning and creep, and did not include the effects associated with the intrinsic structure of the vortex line and the related retardation and inertial effects arising from the dynamics of the electromagnetic field and the superconducting order parameter.
	
	In addition to pinning, the frequency-dependent vortex dynamics also arises from internal effects governed by the effective mass $m$ in the equation of motion \eqref{eq:of:motion:with:m:and:k}. Originally, the vortex mass was introduced phenomenologically by Suhl \cite{SuhlH1965} and then by Gittleman and Rosenblum to model the small-amplitude oscillatory motion of vortices around the pinning centers and to reconcile theory with experiment \cite{GittlemanJI1966}. Subsequently, a variety of approaches of the vortex mass calculation have been developed, ranging from phenomenological models \cite{SuhlH1965} to microscopic theories \cite{KopninNB1978,KopninNB1998,BlatterG1994,OtterloA1995}, in which the mass is related to the inertia of quasiparticle excitations in the vortex core.
	Generally, an adequate analysis of the frequency dependent vortex response should include not only inertial but also retardation effects which can not be described using the mass term in the dynamic equation. Moreover, as demonstrated in \cite{KupriyanovMYu1975}, at temperatures close to the critical temperature $T_{c}$, the vortex mass concept can even lead to some confusion giving the negative sign of the effective mass coefficient (at least within the TDGL theory). Thus, the development of an appropriate theoretical model providing a quantitative description of the intrinsic mechanisms responsible for the inertial and retardation effects in the vortex motion is of primary importance both for the understanding of the physics of vortex matter and for the applications in superconducting cryoelectronics.
	
	Recent microwave studies also emphasize the need for such a theory. Near-field microwave measurements have revealed a nonlinear response of trapped vortices under localized rf excitation \cite{WangCY2025}, while TDGL simulations for vortices in artificial pinning sites have shown a strong nonlinear inductive response \cite{LuhaibiAl2025}. This indicates that the intrinsic dynamic response of a vortex cannot be fully captured by the simple mass model.
	
	It is the goal of our work to fill this gap and derive appropriate equation of vortex motion for a wide range of frequencies. We restrict ourselves by the temperature region close to $T_c$ and start from the simplest version of the model describing the dynamics of the superconducting order parameter, namely, from the TDGL theory.
	
\begin{figure}[!t]
		\includegraphics[width=1.0\linewidth]{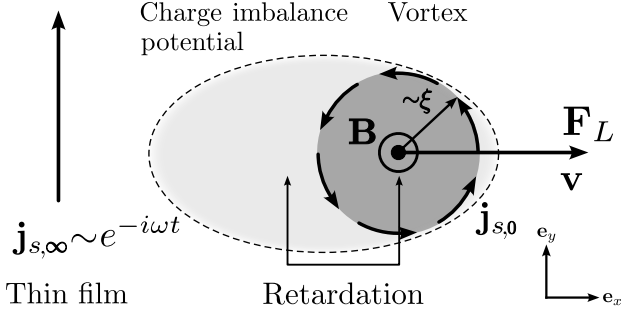}
		\caption{Schematic illustration of a moving vortex with the core radius of the order of coherence length $\xi$ and velocity $\mathbf{v}$. The region with the induced charge-imbalance potential is shown by the dashed line. Here $\mathbf{j}_{s,\infty}$ is the ac transport current producing the Lorentz force $\mathbf{F}_{L}$, $\mathbf{j}_{s,0}$ is superfluid current circulating around the vortex core, $\mathbf{B} = B_{z} \mathbf{e}_{z}$ is magnetic field ($B_{z} > 0$).}\label{pic:sec:ii}
	\end{figure}

	A natural characteristic timescale on which such retardation effects can manifest themselves within the TDGL model is the Ginzburg--Landau relaxation time \cite{TinkhamM2004, KopninNB2001}. Frequency-dependent vortex dynamics with retardation effects can be described by the equation following (see Fig. \ref{pic:sec:ii}):
	\begin{equation}\label{eq-int-aim}
		\left( \eta_{\text{Rel}} + \eta_{\text{Ohm}} \dfrac{1}{1 - i \omega \tau_{E}} \right) \tilde{\mathbf{v}} (\omega) = \tilde{\mathbf{F}}_{L} (\omega),
	\end{equation}
	where $\tilde{\mathbf{F}}_{L} (\omega)$ is the Lorentz force produced by the ac transport current $\tilde{\mathbf{j}}_{s,\infty} (\omega)$ in the frequency representation.
	
	The retarded vortex response resulting from Eq.~\eqref{eq-int-aim} can be explained as follows. The term $\eta_{\text{Ohm}}$ appears due to the conversion of supercurrent into normal current inside the moving vortex core, which occurs at the length scale $l_{E}$. Within the Bardeen-Stephen model with a step like gap profile in the core and $l_{E}^{2} \ll \xi^{2}$ we get $\eta_{\text{Ohm}} \propto l_{E}^{2}/\xi^{2}$ \cite{BardeenJ1965,GenkinVM1989}. The length $l_{E}^{2}$ is proportional to the London penetration depth $\lambda^{2}$. In the nonstationary case the two-fluid model gives us the effective penetration depth $\lambda_{\text{eff}}^{2} (\omega) = \lambda^{2} (1 - i \omega \tau_{E})^{-1}$ \cite{KoganVG2021} and the following relations are preserved: $\eta_{\text{Ohm}} (\omega) \propto l_{E}^{2} (\omega) \propto \lambda_{\text{eff}}^{2} (\omega)$. Thus,
	\begin{equation}
		\dfrac{\eta_{\text{Ohm}}(\omega)}{\eta_{\text{Ohm}}}
		= \dfrac{ l_{E}^{2} (\omega) }{ l_{E}^{2} }
		= \dfrac{\lambda_{\text{eff}}^{2}(\omega)}{\lambda^{2}}
		= \dfrac{1}{1 - i \omega \tau_{E} }.
	\end{equation}
	
	The frequency dependence of the vortex viscosity in Eq.~\eqref{eq-int-aim} leads to qualitatively different behavior in the low- and high-frequency regimes. For $\omega \tau_{E} \ll 1$, Eq.~\eqref{eq-int-aim} contains an effective negative mass $m_{\text{eff}} = - \eta_{\text{Ohm}} \tau_{E}$ (the sign of $m_{\text{eff}}$ is consistent with the result of~\cite{KupriyanovMYu1975}). By contrast, in the regime $\omega \tau_{E} \gg 1$, the Ohmic contribution acts similarly to a pinning term (see Eq.~\eqref{eq:of:motion:with:m:and:k}) with an effective pinning constant $\eta_{\mathrm{Ohm}}/\tau_{E}$. This results in a renormalization of the depinning frequency $\omega_k$, while the vortex viscosity becomes purely relaxational.
	
	The aim of this work is to derive Eq.~\eqref{eq-int-aim} within the framework of the TDGL model describing gapless superconductivity and to investigate how the retarded vortex dynamics affects the parameters of superconducting circuits, e.g., impedance or kinetic inductance. 
	
	The paper is organized as follows. In Section~\ref{sec:ii}, we present the basic equations of the TDGL model. In Section~\ref{sec:iii}, we derive the equation of motion for a vortex. In Section~\ref{sec:iv}, we analyze the corrections to the kinetic inductance arising from this nonstationary vortex dynamics. In Section~\ref{sec:v}, we summarize  our results.
	
	\section{MODEL AND BASIC EQUATIONS}\label{sec:ii}
	
	{As noted in the} introduction{,} our study of the vortex dynamics will be {based} on the time-dependent Ginzburg--Landau (TDGL) theory \cite{TinkhamM2004, KopninNB2001}:
	\begin{equation}\label{eq:TDGL:psi}
		\dfrac{1}{D}
		\left( 
		\dfrac{\partial \psi}{\partial t} - i \dfrac{2 \pi c}{\Phi_{0}} \phi \psi
		\right) =
		(\nabla + i \dfrac{2 \pi}{\Phi_{0}} \mathbf{A} )^{2} \psi + 
		\xi^{-2} \left( 1 - \dfrac{|\psi|^{2}}{|\psi_{\infty}|^{2}} \right) \psi,
	\end{equation}
	\begin{equation}\label{eq:TDGL:j}
		\mathbf{j} = \mathbf{j}_{n} + \mathbf{j}_{s} = \sigma \mathbf{E} - \dfrac{c}{4 \pi \lambda^{2}} \dfrac{|\psi|^{2}}{|\psi_{\infty}|^{2}} \left( \mathbf{A} + \dfrac{\Phi_{0}}{2 \pi} \nabla \chi \right),
	\end{equation}
	where $D = l v_{F} / 3$ is the diffusion coefficient of {the} normal metal, {$ \phi$ is the scalar potential,} $\xi^{2} = \pi \hbar D / \bigl(8 (T_{c} - T)\bigr)$ is the coherence length,  $\lambda^{2} = \hbar c^{2} T_{{c}} / ( 2 \sigma \pi^{2} |\psi_{\infty}|^{2} )$ is the magnetic penetration depth, $|\psi_{\infty}|^{2} = 8 \pi^{2} T_{{c}} (T_{c} - T) / (7 \zeta (3) )$ is the equilibrium value of the order parameter, $\Phi_{0} = \pi \hbar c / |e|$ is the superconducting flux quantum, {$\sigma$ is the dc normal-state conductivity evaluated at the onset of the superconducting transition $T\simeq T_c$ ($\sigma$ is treated as frequency and temperature independent).}.
	
	{The above TDGL model applies to gapless superconductors and can be derived microscopically in two limiting cases: (i) in the presence of a high concentration of magnetic impurities ($u^{2} = \xi^{2} / l_{E}^{2} = 12$) \cite{GorkovLP1968,KopninNB2001} and (ii) when inelastic electron-phonon relaxation is taken into account ($u^{2} = \xi^{2} / l_{E}^{2} = 5.79$) \cite{KramerL1978,WattsTobinRJ1981}. In the latter case, the validity range of the TDGL theory is given by the conditions
		\begin{equation}
			\omega \tau_{ph} \ll 1, \qquad 1 - \dfrac{T}{T_{c}} \ll \frac{\hbar}{\tau_{ph} T_c},
		\end{equation}
		where $\tau_{ph}$ is the inelastic electron-phonon scattering time, $\omega$ is the frequency of the ac transport current $\mathbf{j}_{s,\infty}$ and $l_{E}$ is the spatial scale for conversion of the supercurrent $\mathbf{j}_{s}$ into normal current $\mathbf{j}_{n}$. Although the TDGL model is  formally derived only for gapless superconductors in the dirty regime, it can still provide a qualitative understanding of vortex dynamics even beyond its strict range of applicability.
	}
	
	The frequency {$\omega$} of the ac transport current $\mathbf{j}_{s,\infty}$ is assumed to be smaller than the plasma frequency of the electronic system, therefore, neglecting the time derivative of the charge density in the continuity equation, {one} can obtain \cite{GorkovLP1975}
	\begin{equation}\label{eq:nabla:j}
		\nabla \mathbf{j} = 0.
	\end{equation}
	
	In the limit $\varkappa=\lambda/\xi \gg 1$, the vector potential generated by the vortex screening currents is negligibly small at the length scales $r \ll \lambda$ and therefore the gauge-invariant combination $\mathbf{A} + (\Phi_{0}/2\pi)\nabla\chi$ reduces to $(\Phi_{0}/2\pi)\nabla\chi$. We choose a gauge in which $\mathbf{j}_{s,\infty}$ is represented by a spatially uniform vector potential $\mathbf{A}$, yielding
	\begin{equation}\label{eq:js:infty:A}
		\mathbf{j}_{s,\infty}(t)=-\frac{c}{4\pi\lambda^{2}}\mathbf{A}(t).
	\end{equation}
	
	{The above limit $\varkappa^{2} \gg 1$ (strong type-II superconductor) is consistent with condition $u^{2} = \xi^{2} / l_{E}^{2} \gg 1$, which we will use below. Indeed the value $l_{E}$ is related to the magnetic penetration depth $\lambda$ as follows:
		\begin{equation}\label{eq:l_E}
			l_{E}^{2} = \frac{4 \pi D \sigma}{c^{2}} \lambda^{2}.
		\end{equation}
		Dividing Eq.~\eqref{eq:l_E} by the $\xi^{2}$ and taking $D \sim 1 \text{ cm}^{2}/\text{s}$ and $\sigma \sim 10^{17} \text{ s}^{-1}$ (which corresponds to $10 \text{ MS}/\text{m}$), one obtains $\varkappa^{2} u^{2} \sim 10^{3}$. Thus, the limit $\varkappa^{2} \gg 1$ is fully consistent with the condition $u^{2} \gg 1$.
	}
	
	\section{EQUATION OF MOTION FOR SINGLE ISOLATED VORTEX}\label{sec:iii}
	
	To derive the equation of motion for the vortex, we employ perturbation theory, assuming that the position of the vortex is defined by the vector $\mathbf{R}(t)$ and that the vortex moves slowly with velocity $\mathbf{v} = \partial_{t} \mathbf{R}$. We choose $\psi$ in the following form (in the literature, this approach is often referred to as the collective–coordinate method):
	\begin{equation}\label{eq:psi0:psi1}
		\psi \approx \psi_{0} + \psi_{1}
		\approx |\psi_{\infty}| (f_{0} + f_{1}) e^{i \chi_{0} + i \chi_{1}}.
	\end{equation}
	Here $|\psi_{\infty}| f_{0}$ and $\chi_{0}$ are the magnitude and phase of the order parameter corresponding to a static vortex shifted by $\mathbf{R}(t)$. The corrections $|\psi_{\infty}| f_{1} (t, \mathbf{r} - \mathbf{R}(t))$ and $\chi_{1}(t, \mathbf{r} - \mathbf{R}(t))$ are assumed to be small and of the first order in the vortex velocity $\mathbf{v}$. It is important to note that the functions $f_{1}$ and $\chi_{1}$ depend not only on the moving coordinate $\mathbf{r} - \mathbf{R}(t)$, but also explicitly contain time $t$.
	
	To simplify the problem, we transform to a reference frame moving with the vortex center: 
	\begin{equation}\label{eq:moving:frame}
		\mathbf{r} \rightarrow \mathbf{r} + \mathbf{R}(t), \qquad \partial_{t} \rightarrow \partial_{t} - (\mathbf{v}, \nabla).
	\end{equation}
	In this frame $f_{0} = f_{0} (\rho)$ and $\chi_{0} = - \varphi$, where $\rho$ is the radial distance and $\varphi$ is the polar angle in a cylindrical coordinate system with the polar axis aligned with $\mathbf{e}_{x}$ in Fig. \ref{pic:sec:ii}). The function $f_{0}$ is the solution of the following equation:
	\begin{equation}\label{eq:f0}
		(\nabla^{2} - \rho^{-2}) f_{0} + \xi^{-2} (1 - f_{0}^{2}) f_{0} = 0.
	\end{equation}
	
	Upon expanding Eq.$\,$\eqref{eq:TDGL:psi} to first order in $\mathbf{v}$, the nonlinear term $|\psi|^{2}\psi$ takes the form $|\psi_{0}|^{2}\psi_{0} + 2 |\psi_{0}|^{2}\psi_{1} + \psi_{0}^{2}\psi_{1}^{*}$. Thus, in first order in perturbation theory we obtain a matrix equation for the column vector $\begin{pmatrix} \psi_{1}, & \!\! \psi_{1}^{*} \end{pmatrix}^{T}$:
	\begin{equation}\label{eq:psi1}
		- \frac{1}{D} \frac{\partial}{\partial t}
		\begin{pmatrix}
			\psi_{1} \\ \psi_{1}^{*}
		\end{pmatrix}
		=
		\hat{H}_{1}
		\begin{pmatrix}
			\psi_{1} \\ \psi_{1}^{*}
		\end{pmatrix}
		+ S,
	\end{equation}
	where the Hamiltonian $\hat{H}_{1}$ and the source term $S$ are given by
	\begin{equation}\label{eq:H1}
		\hat{H}_{1} =
		\begin{pmatrix}
			- \nabla^{2} + \xi^{-2} (2 f_{0}^{2} - 1)
			& \xi^{-2} \dfrac{\psi_{0}^{2}}{|\psi_{\infty}|^{2}} \\[4pt]
			\xi^{-2} \dfrac{\psi_{0}^{*2}}{|\psi_{\infty}|^{2}}
			& - \nabla^{2} + \xi^{-2} (2 f_{0}^{2} - 1)
		\end{pmatrix},
	\end{equation}
	\begin{equation}\label{eq:S}
		S =
		\begin{pmatrix}
			\left( - \dfrac{1}{D} \mathbf{v} + \dfrac{4 \pi i}{\Phi_{0}} \mathbf{A}, \nabla \right) \psi_{0}
			+ \dfrac{i}{D} \psi_{0} \left( \dfrac{\partial \chi_{1}}{\partial t} - \dfrac{2 \pi c}{\Phi_{0}} \phi \right)
			\\[10pt]
			\left( - \dfrac{1}{D} \mathbf{v} - \dfrac{4 \pi i}{\Phi_{0}} \mathbf{A}, \nabla \right) \psi_{0}^{*}
			- \dfrac{i}{D} \psi_{0}^{*} \left( \dfrac{\partial \chi_{1}}{\partial t} - \dfrac{2 \pi c}{\Phi_{0}} \phi \right)
		\end{pmatrix}.
	\end{equation}
	
	The function $(\mathbf{d},\nabla) \begin{pmatrix} \psi_{0}, & \!\! \psi_{0}^{*} \end{pmatrix}^{T}$, where $\mathbf{d}$ is an arbitrary constant vector, is an eigenfunction of the Hamiltonian $\hat{H}_{1}$ with zero eigenvalue (see Eq.~\eqref{eq:f0}). Therefore, multiplying Eq.$\,$\eqref{eq:psi1} from the left by $(\mathbf{d}, \nabla) \begin{pmatrix} \psi_{0}^{*}, & \!\! \psi_{0} \end{pmatrix}$ and integrating over the film area, we obtain the following integral relation (integration along the $z$ axis is omitted since the film is thin):
	\begin{multline}\label{eq:int:cond}
		\frac{1}{2 \pi |\psi_{\infty}|^{2}}
		\int d^{2} \mathbf{r}\,
		(\mathbf{d} {\,\cdot\,} \nabla)
		\begin{pmatrix}
			\psi_{0}^{*}, & \!\! \psi_{0}
		\end{pmatrix}
		\frac{\partial}{\partial t}
		\begin{pmatrix}
			\psi_{1} \\ \psi_{1}^{*}
		\end{pmatrix}
		= \\
		= - \frac{D}{2 \pi |\psi_{\infty}|^{2}}
		\int d^{2} \mathbf{r}\,
		(\mathbf{d} {\,\cdot\,} \nabla)
		\begin{pmatrix}
			\psi_{0}^{*}, & \!\! \psi_{0}
		\end{pmatrix}
		S.
	\end{multline}
	The integral on the right-hand side of Eq.$\,$\eqref{eq:int:cond}, with Eq.$\,$\eqref{eq:js:infty:A} taken into account, takes the form:
	\begin{multline}\label{eq:S:mod}
		- \frac{D}{2 \pi |\psi_{\infty}|^{2}}
		\int d^{2} \mathbf{r}\,
		(\mathbf{d}, \nabla)
		\begin{pmatrix}
			\psi_{0}^{*}, & \!\! \psi_{0}
		\end{pmatrix}
		S
		= \\
		(\mathbf{d}, \mathbf{v})
		\int \rho\, d\rho
		\left( \frac{\partial f_{0}}{\partial \rho} \right)^{2}
		- \frac{16 \pi^{2} D \lambda^{2}}{\Phi_{0} c}
		\bigl(\mathbf{d},
		[\,\mathbf{j}_{s,\infty} {\,\times\,} \mathbf{e}_{z}]\bigr) +
		\\
		+ \frac{1}{\pi}
		\int d^{2} \mathbf{r}\,
		(\mathbf{d}, \mathbf{e}_{\varphi})
		\frac{f_{0}^{2}}{\rho}
		\left[
		(\mathbf{v}, \mathbf{e}_{\varphi}) \frac{1}{\rho}
		+ \frac{\partial \chi_{1}}{\partial t}
		- \frac{2 \pi c}{\Phi_{0}} \phi
		\right].
	\end{multline}
	The integrals in the right-hand side of Eq.~\eqref{eq:S:mod} were evaluated in the stationary case, i.e., for a dc current $\mathbf{j}_{s,\infty}$, in works \cite{KupriyanovMYu1972,HuCR1972,HuCR1973} under the condition $\xi^{2} / l_{E}^{2} = u^{2} = 12$:
	\begin{equation}\label{eq:int:f}
		(\mathbf{d}, \mathbf{v}) \int \rho\, d\rho
		\left( \frac{\partial f_{0}}{\partial \rho} \right)^{2}
		= \alpha_{\text{Rel}}\, (\mathbf{d}, \mathbf{v})
		\approx 0.279\, (\mathbf{d}, \mathbf{v}),
	\end{equation}
	\begin{multline}\label{eq:int:mu}
		\frac{1}{\pi} \int d^{2} \mathbf{r}\,
		(\mathbf{d}, \mathbf{e}_{\varphi})
		\frac{f_{0}^{2}}{\rho}
		\left[
		(\mathbf{v}, \mathbf{e}_{\varphi})\frac{1}{\rho}
		- \frac{2 \pi c}{\Phi_{0}} \phi
		\right]
		= \\
		= \frac{1}{\pi} \int d^{2} \mathbf{r}\,
		(\mathbf{d}, \mathbf{e}_{\varphi})
		\frac{f_{0}^{2}}{\rho}\, \mu
		= \alpha_{\text{Ohm}}\, (\mathbf{d}, \mathbf{v})
		\approx 0.159\, (\mathbf{d}, \mathbf{v}),
	\end{multline}
	where {$\mu$ is the charge-imbalance potential:
		\begin{equation}\label{eq:mu}
			\mu = \frac{\partial \chi}{\partial t}
			- \frac{2 \pi c}{\Phi_{0}} \phi
			\approx
			(\mathbf{v}, \mathbf{e}_{\varphi})\frac{1}{\rho}
			+ \frac{\partial \chi_{1}}{\partial t}
			- \frac{2 \pi c}{\Phi_{0}} \phi.
	\end{equation}}
	
	The coefficients $\alpha_{\text{Rel}}$ and $\alpha_{\text{Ohm}}$ can be interpreted as the dimensionless relaxational and Ohmic viscosities, respectively, associated with order-parameter relaxation and with Ohmic dissipation due to vortex motion. It is important to note that, in the nonstationary case, the integral in Eq.~\eqref{eq:int:mu} additionally includes the term $\partial_{t} \chi_{1}$, because $\partial_{t} \chi_{1}$ is of order $\omega \mathbf{v}$ (where $\omega$ is the frequency of $\mathbf{j}_{s,\infty}$), unlike in the stationary vortex motion, where it is of order $\mathbf{v}^{2}$.
	
	In the time-dependent case it is convenient to evaluate the integral Eq.~\eqref{eq:int:mu} using a Fourier transform, which for an arbitrary function
	$F(t,\mathbf{r})$ is written as:
	\begin{equation}\label{eq:Fourier}
		F(t,\mathbf{r}) = \int \frac{d \omega}{2 \pi}
		e^{-i \omega t} \tilde{F} (\omega, \mathbf{r}),
		\qquad
		\partial_{t} \rightarrow - i \omega.
	\end{equation}
	Then, in the limit $\xi^{2} / l_{E}^{2} = u^{2} \gg 1$, the integral Eq.$\,$\eqref{eq:int:mu} takes the form:
	\begin{equation}\label{eq:int:mu:t}
		\frac{1}{\pi} \int d^{2} \mathbf{r}\,
		(\mathbf{d}, \mathbf{e}_{\varphi})
		\frac{f_{0}^{2}}{\rho}\, \tilde{\mu}
		= \frac{\alpha_{\text{Ohm}}}{1 - i \omega \tau_{E}}
		(\mathbf{d}, \mathbf{\tilde{v}}),
	\end{equation}
	where $\alpha_{\text{Ohm}} \approx 2 u^{-2}$ and
	$\tau_{E} = l_{E}^{2} / D \propto (T_{c} - T)^{-1}$ is the characteristic timescale of variation of the gauge-invariant potential $\mu$, i.e., the Ginzburg--Landau time. The corresponding calculation yielding Eq.~\eqref{eq:int:mu:t} is given in Appendix~\ref{Ap:A}. Combining equations$\,$\eqref{eq:int:f}, \eqref{eq:Fourier}, and \eqref{eq:int:mu:t},
	we obtain from Eq.~\eqref{eq:S}:
	\begin{multline}\label{eq:v:int:cond}
		- \frac{i \omega}{2 \pi |\psi_{\infty}|^{2}}
		\int d^{2} \mathbf{r}\,
		(\mathbf{d}, \nabla)
		\begin{pmatrix}
			\psi_{0}^{*}, & \!\! \psi_{0}
		\end{pmatrix}
		\begin{pmatrix}
			\tilde{\psi}_{1} \\ \tilde{\psi}_{1}^{*}
		\end{pmatrix}
		= \\
		= \left(
		\alpha_{\text{Rel}}
		+ \frac{\alpha_{\text{Ohm}}}{1 - i \omega \tau_{E}}
		\right)
		(\mathbf{d}, \mathbf{\tilde{v}})
		- \frac{16 \pi^{2} D \lambda^{2}}{\Phi_{0} c}
		(\mathbf{d}, [\,\mathbf{\tilde{j}}_{s, \infty} {\,\times\,} \mathbf{e}_{z}]).
	\end{multline}
	The left-hand side of Eq.~\eqref{eq:v:int:cond} vanishes due to the solvability condition associated with the translational zero mode of the vortex. Indeed, the correction $\psi_{1}$ is defined only up to an additive contribution proportional to $(\mathbf{d}, \nabla) \psi_{0}$ which is equivalent to an infinitesimal shift of the vortex position. Fixing the vortex coordinate therefore requires the orthogonality condition:
	\begin{equation}
		\int d^{2} \mathbf{r} (\mathbf{d}, \nabla)
		\begin{pmatrix}
			\psi_{0}^{*}, & \!\! \psi_{0}
		\end{pmatrix}
		\begin{pmatrix}
			\tilde{\psi}_{1} \\ \tilde{\psi}_{1}^{*}
		\end{pmatrix} = 0.
	\end{equation}
	In the Fourier representation, this gives the left-hand side of Eq.~\eqref{eq:v:int:cond} and hence it is identically zero.
	
	Thus, dropping an arbitrary vector $\mathbf{d}$ from the scalar products in Eq.~\eqref{eq:v:int:cond} and including the pinning forces, we obtain the equation of motion of a vortex line driven by the Lorentz force $\mathbf{\tilde{F}}_{L} (\omega) = (\Phi_{0} / c)\,[\,\mathbf{\tilde{j}}_{s,\infty} (\omega) {\,\times\,} \mathbf{e}_{z}]$:
	\begin{equation}\label{eq:v}
		\eta(\omega)\,\mathbf{\tilde{v}} = \mathbf{\tilde{F}}_{L} (\omega),
	\end{equation}
	where
	\begin{equation}\label{eq:eta}
		\eta(\omega) = \eta_{\text{Rel}} + \eta_{\text{Ohm}} \dfrac{1}{1 - i \omega \tau_{E}} - \eta_{\text{tot}} \dfrac{\omega_{k}}{i \omega}
	\end{equation}
	and $\eta_{\text{tot}} = \eta_{\text{Rel}} + \eta_{\text{Ohm}}$. The relaxational and Ohmic viscosities are
	\begin{equation}\label{eq:eta:rel:ohm}
		\eta_{\text{Rel}} = \frac{\sigma \Phi_{0} H_{c2}}{c^{2}} \dfrac{ \alpha_{\text{Rel}} u^{2}}{2} , \quad \eta_{\text{Ohm}} = \frac{\sigma \Phi_{0} H_{c2}}{c^{2}} \dfrac{\alpha_{\text{Ohm}} u^{2}}{2},
	\end{equation}
	where $H_{c2} = \Phi_{0} / (2 \pi \xi^{2})$ is the upper critical field \cite{KopninNB2001}. The term $\eta_{tot} \omega_{k}/(- i \omega)$ corresponds to the linear pinning force in the frequency domain. The depinning frequency $\omega_{k}$ depends strongly on the type of the pinning centers. For a vortex pinned by a cavity or an insulating inclusion~\cite{BespalovAA2013}, one obtains, e.g.:
	\begin{equation}\label{eq:k}
		\omega_{k} \tau_{E} = \dfrac{2}{\alpha u^{2}} \dfrac{3 f'(0)^{2} a^{2}}{2},
	\end{equation}
	where $f'(0) \approx 0.583 / \xi$, $a$ is the characteristic length scale of the pinning center and $\alpha = \alpha_{\text{Rel}} + \alpha_{\text{Ohm}}$.
	
	The analysis of Eq.~\eqref{eq:v} leads to the following conclusions. (i) First, a linear expansion of the kernel $(1 - i \omega \tau_{E})^{-1}$ in powers of frequency yields a negative effective mass $m_{\text{eff}} = - \eta_{\text{Ohm}} \tau_{E}$:
	\begin{equation}
		\mathbf{F}_{L}
		\approx (\eta_{\text{Rel}} + \eta_{\text{Ohm}})\,\mathbf{v}
		- \eta_{\text{Ohm}} \tau_{E}\,\dot{\mathbf{v}}.
	\end{equation}
	Therefore, a correct description of the temporal vortex dynamics requires the full kernel $(1 - i \omega \tau_{E})^{-1}$ rather than a Taylor expansion of Eq.$\,$\eqref{eq:v} in $\omega$. Moreover, the sign of $m_{\text{eff}}$ is consistent with the numerical results of \cite{KupriyanovMYu1975}. (ii) Second, the form of the kernel indicates that the vortex dynamics exhibits a retardation effect with a characteristic timescale $\tau_{E}$. (iii) Third, at high frequencies, $\omega \tau_{E} \gg 1$, the kernel becomes $-1/(i \omega \tau_{E})$. This form is similar to the contribution from pinning forces in Eq.~\eqref{eq:eta}. In the time domain, it corresponds to a restoring elastic force arising from the backaction of the charge-imbalance potential $\mu$, induced by the vortex motion.
	
	{Note, that the amplitude of vortex oscillations becomes smaller in the limit $\omega \tau_{E} \gg 1$ which may reduce the vortex contribution to the experimentally measurable quantities. 
		This amplitude of vortex oscillations can be of course increased for higher
		transport current density. Note, however, that increasing the transport current we can  simultaneously increase the nonlinear corrections to Eq.~\eqref{eq:v} which have the order $( |\mathbf{j}_{s,\infty}|/j_{\text{crit}} )^{3}$, where $j_{\text{crit}}$ is the depairing critical current. These corrections can hamper the experimental observation of the predicted pinning like behavior  since these corrections will affect the characteristics of resonators used for impedance measurements.
		Still for materials with rather high depairing currents the operating window for the linear response regime could be experimentally accessible.
	}
	
	\section{VORTEX CONTRIBUTION TO MICROWAVE IMPEDANCE: ROLE OF CHARGE-IMBALANCE POTENTIAL AFFECTED IN THE VORTEX DYNAMICS}\label{sec:iv}
	
	In this section, we derive the vortex-related contribution to the kinetic inductance of a superconductor, explicitly taking into account the effect of nonstationary vortex dynamics. For an applied ac transport current, the spatially averaged induced electric field can be written as (see Appendix~\ref{Ap:B})
	\begin{equation}\label{eq:induced:E}
		- i \omega \langle\, \tilde{\mathbf{j}}_{s} \rangle = \dfrac{c^{2}}{4 \pi \lambda^{2}} \left( \langle \tilde{\mathbf{E}} \rangle + \dfrac{1}{c} [\tilde{\mathbf{v}}, \langle \mathbf{B}_{v} \rangle] \right),
	\end{equation}
	where $\langle\, \tilde{\mathbf{j}}_{s} \rangle$ is the spatially averaged supercurrent density, and $\langle \mathbf{B}_{v} \rangle$ includes the magnetic field associated with vortices. In the dc limit, $\omega = 0$, Eq.~\eqref{eq:induced:E} reduces to the standard relation $ \langle \mathbf{E} \rangle = - \dfrac{1}{c} [\mathbf{v}, \langle \mathbf{B}_{v} \rangle]$ as expected for steady vortex motion~\cite{KopninNB2001}. In the absence of vortices, $\langle \mathbf{B}_{v} \rangle = 0$, Eq.~\eqref{eq:induced:E} reduces to the London equation~\cite{TinkhamM2004}.
	
	Using Eq.~\eqref{eq:v}, we obtain
	\begin{equation}
		- i \omega \dfrac{4 \pi \lambda^{2}}{c^{2}} \langle\, \tilde{\mathbf{j}}_{s} \rangle + \dfrac{\Phi_{0} |\langle \mathbf{B}_{v} \rangle|}{c^{2} \eta(\omega)} \mathbf{\tilde{j}}_{s,\infty} = \langle \tilde{\mathbf{E}} \rangle.
	\end{equation}
	
	A quantitative evaluation of the currents $\langle \tilde{\mathbf{j}}_{s} \rangle$ and $\mathbf{\tilde{j}}_{s,\infty}$ would, in general, require a self-consistent solution of the Maxwell equations, taking into account the sample geometry, the electrical boundary conditions imposed by the current-injection scheme (e.g., capacitive coupling), and the time-dependent spatial distribution of vortices. Such a treatment is beyond the scope of the present discussion. Therefore, in what follows, we restrict ourselves to the analysis of the vortex-related contribution $z_{v}$ to the total impedance of the superconductor $(z_{0} + z_{v}) \mathbf{\tilde{j}}_{\text{tr}} = \langle \mathbf{\tilde{E}} \rangle$, where $z_{0}$ is the impedance without vortices. Within the simplifying approximation $\mathbf{\tilde{j}}_{s,\infty} = \mathbf{\tilde{j}}_{\text{tr}}$ (although, strictly speaking, the transport current at finite frequency also contains a normal component), one finds
	\begin{equation}\label{eq:vortex:z}
		z_{v} = \dfrac{1}{\sigma} \dfrac{2}{\alpha u^{2}} \dfrac{| \langle \mathbf{B}_{v} \rangle |}{ H_{c2} } \dfrac{1}{ \dfrac{\alpha_{\text{Rel}}}{\alpha} + \dfrac{\alpha_{\text{Ohm}}}{\alpha} \dfrac{1}{ 1 - i \omega \tau_{E} } - \dfrac{\omega_{k}}{i \omega} }.
	\end{equation}
	
	In the low-frequency limit and for weak pinning, $\omega \tau_{E} \ll 1$ and $\omega_{k} \ll \omega$, Eq.~\eqref{eq:vortex:z} reduces to
	\begin{equation}
		z_{v} = \dfrac{1}{\sigma} \dfrac{2}{\alpha u^{2}} \dfrac{| \langle \mathbf{B}_{v} \rangle |}{ H_{c2} } \left( \dfrac{\omega_{k}}{i \omega} + 1 - i \dfrac{ \alpha_{\text{Ohm}} }{ \alpha } \omega \tau_{E} \right).
	\end{equation}
	The first term, $\omega_{k}/(i\omega)$, describes the reactive contribution induced by pinning. The second term, equal to $1$, corresponds to the conventional flux-flow resistivity. The last term $-i(\alpha_{\text{Ohm}} / \alpha) \omega \tau_{E}$ represents an intrinsic kinetic-inductive correction originating from the relaxation of the charge-imbalance potential. The corresponding correction to the kinetic inductance $l_{k,v}$ is given by
	\begin{equation}\label{eq:kin:ind:vortex}
		l_{k,v} = l_{k} \dfrac{2}{\alpha u^{2}} \dfrac{| \langle \mathbf{B}_{v} \rangle |}{H_{c2}} \dfrac{\alpha_{\text{Ohm}}}{\alpha},
	\end{equation}
	where $l_{k} = 4 \pi \lambda^{2} / c^{2}$. By contrast, in the opposite limit of sufficiently strong pinning, $\omega \tau_{E} \ll 1$ and $\omega_{k} \gg \omega$, the impedance $z_{v}$ reduces to the standard expression (see, e.g. \cite{CoffeyMW1991}).
	
	In the high-frequency regime, $\omega \tau_{E} \gg 1$, the vortex-related impedance takes a form similar to that the one obtained in Ref.~\cite{CoffeyMW1991}:
	\begin{equation}
		z_{v} = \dfrac{1}{\sigma} \dfrac{2}{\alpha_{\text{Rel}} u^{2}} \dfrac{| \langle \mathbf{B}_{v} \rangle |}{ H_{c2} } \dfrac{1}{ 1 - \dfrac{\omega_{k}'}{i \omega}  },
	\end{equation}
	where
	\begin{equation}\label{eq:omega:k:renorm}
		\omega_{k}' = \dfrac{\alpha_{\text{Ohm}}}{\alpha_{\text{Rel}}} \tau_{E}^{-1} +\dfrac{\alpha}{\alpha_{\text{Rel}}} \omega_{k}
	\end{equation}
	is the renormalized depinning frequency. In this limit, the vortex viscosity becomes purely relaxational:
	\begin{equation}\label{eq:eta:tot:renorm}
		\eta_{\text{tot}}' = \dfrac{ \alpha_{\text{Rel}} }{\alpha} \eta_{\text{tot}} = \eta_{\text{Rel}}.
	\end{equation}
	Accordingly, the nonequilibrium Ohmic channel no longer enters the viscosity directly, but instead manifests itself through the renormalization of the depinning frequency.
	
	The operating frequencies in our consideration can not exceed  the microscopic relaxation scales entering the TDGL derivation, i.e., the electron-phonon relaxation rate $\tau_{ph}^{-1}$. Thus, the high-frequency regime discussed above is defined according to the
	conditions
	\begin{equation}
		\tau_E^{-1}\ll \omega \ll \tau_{ph}^{-1}
	\end{equation}
	provided that such frequency window can exist. For a fully gapped superconductor one should also take account of an additional pair-breaking threshold, roughly $\hbar\omega\sim 2\Delta(T)$, which can also affect affect  the dynamics of the charge imbalance potential. This regime is, however, outside the gapless TDGL model used in our work and requires a separate  treatment on the basis of more elaborate microscopic theory.

	It is worth emphasizing that, within the TDGL framework adopted in Sec.~\ref{sec:ii}, the characteristic retardation time $\tau_{E}$, as well as the associated length scale $l_{E}$ and parameter $u^{2}=\xi^{2}/l_{E}^{2}$, are not universal quantities but depend on the microscopic assumptions underlying the derivation of the effective equations. For the specific gapless TDGL model defined by Eqs.~\eqref{eq:TDGL:psi} and \eqref{eq:TDGL:j}, one obtains the estimate
	\begin{equation}
		\tau_{E} = \dfrac{\tau_{E}(0)}{1 - (T/T_{c})}, \quad \tau_{E}^{-1}(0) \sim \dfrac{T_{c}}{\hbar}
	\end{equation}
	and $\tau_{E}^{-1} \sim 10^{10}$ Hz for $T/T_{c} = 0.9$ and $T_{c} / k_{B} = 1$ K. {Note that for a generalized TDGL model (see, e.g., \cite{KopninNB2001,GulianAM2002})  the quantity $\tau_E$ can also depend on the order parameter amplitude $|\psi_{\infty}|$ and electron-phonon relaxation time $\tau_{ph}$.} 
	
	At the same time, in many experimentally relevant superconducting systems one can expect the opposite hierarchy $l_{E} \gg \xi$, in contrast to the limit $\xi \,\,{\gtrsim}\,\, l_{E}$ used in {derivation of the above TDGL model}. In this regime, the spatial distributions of the electric field $\mathbf{E}$ and the gauge-invariant potential $\mu$ extend well beyond the vortex core, which can substantially modify the Ohmic contribution to the vortex viscosity and increase both $\eta_{\mathrm{Ohm}}$ and the retardation time $\tau_{E}$. As a result, the vortex-induced reactive contribution to the impedance, Eq.~\eqref{eq:vortex:z}, and the corresponding correction to the kinetic inductance, Eq.~\eqref{eq:kin:ind:vortex}, may be significantly enhanced. {A q}uantitative description of this regime requires going beyond the short-$l_{E}$ approximation employed here and solving equations for $\mu$ and $\mathbf{E}$ without assuming $l_{E} \ll \xi$.
	
	\section{CONCLUSIONS}\label{sec:v}
	
	To sum up, within the time-dependent Ginzburg--Landau framework for gapless superconductors, we have developed a description of ac-driven vortex dynamics that incorporates relaxation of the gauge-invariant charge-imbalance potential {in the linear response regime}. The main result is the vortex equation of motion, Eqs.~\eqref{eq:v} and \eqref{eq:eta}, in which the Ohmic part of the viscosity is governed by the retarded factor $(1-i\omega\tau_{E})^{-1}$.
	
	It is worth noting that a similar retardation mechanism may also work in superconductors with a more complex structure of the order parameter (e.g., multicomponent order parameter \cite{SilaevM2016} or nematic order parameter \cite{CastilloMF2025}). Certainly in this case the equation for the charge-imbalance potential $\mu$ \eqref{eq:app:A:mu:closed:form} should be modified to account for a more complex form of the superfluid order parameter, however, the kernel should keep the form $(1 - i \omega \tau)^{-1}$ with an appropriate modification of the relaxation time $\tau$. A full derivation for such systems is beyond the scope of the present work.
	
	Our analysis revealed two possible manifestations of this retardation effect. In the low-frequency regime, an expansion of Eq.~\eqref{eq:eta} yields a negative effective mass. In the opposite limit, $\omega\tau_{E}\gg 1$, the retarded Ohmic contribution becomes predominantly reactive and acts as an additional pinning term, renormalizing the depinning frequency. Thus, the relaxation of charge-imbalance potential affects vortex dynamics in two ways: (i) through the dissipative and inertial response at low frequencies and (ii) through the reactive response at high frequencies.
	
	We also derived the vortex contribution to the complex impedance and the corresponding correction to the kinetic inductance. In particular, in the weak-pinning and low-frequency regime, relaxation of charge-imbalance potential produces a reactive correction to the vortex impedance in addition to the standard flux-flow term, while in the high-frequency regime it contributes to the renormalization to the depinning frequency, with the dissipative component of the response being governed by the relaxational viscosity.
	
	Note finally that in the present work we considered only the low magnetic field limit well below   the upper critical field $H_{c2}$. Certainly, for larger magnetic fields $H$  the regions with the induced charge-imbalance potential can overlap and this overlap  can affect the relaxation dynamics and invalidate the isolated-vortex approximation. The derivation of TDGL model assumes that the  $l_E$ length scale is of the order or less than the coherence length (i.e., the vortex core size) and, thus, direct overlap of $l_E$-sized charge-imbalance regions can occur only near $H_{c2}$. In experimentally relevant regimes with $l_E\gtrsim\xi$, however, charge-imbalance regions may overlap already when the intervortex distance becomes comparable to $l_E$. This regime is outside the short-$l_E$ approximation and requires solving the equation for charge imbalance potential for a vortex lattice.

	In both cases, the reduction of the characteristic order-parameter amplitude between vortices can increase the effective relaxation time. Indeed, at fields close to $H_{c2}$ the normalized order-parameter amplitude is reduced. In a rough estimate one may replace
	\begin{equation*}
		\tau_E\rightarrow \frac{\tau_E}{f_{0,H}^2},
	\end{equation*}
	where $f_{0,H}$ is the characteristic normalized value of the order parameter between vortex centers. Consequently, retardation effects may appear at lower frequencies in a dense vortex lattice.

	\section*{ACKNOWLEDGMENTS}
	
	This work was supported by the Russian Science Foundation (Grant No. 25-12-00042) for part of the derivation of the vortex equation of motion and by the Grant of the Ministry of Science and Higher Education of the Russian Federation No. 075-15-2025-010 for part of the calculation of the vortex contribution to the sample impedance.

	\appendix
	\renewcommand{\thesection}{\Alph{section}}
	
	\titleformat{\section}[block]
	{\normalfont\normalsize\bfseries} 
	{Appendix~\thesection:}{1em}{}
	
	\numberwithin{equation}{section}
	\renewcommand{\theequation}{\thesection\arabic{equation}}
	
	\section{Contribution of frequency dependent charge imbalance potential relaxation in the vortex dynamics.}\label{Ap:A}
	
	To calculate integral \eqref{eq:int:mu:t}, we need a closed equation for $\mu$. To obtain it, we take the imaginary part of Eq.~\eqref{eq:TDGL:psi} and multiply it by $|\psi|$:
	\begin{equation}\label{eq:app:A:TDGL:psi:Im}
		\dfrac{|\psi|^{2}}{D} \left( \dfrac{\partial \chi}{\partial t} - \dfrac{2 \pi c}{\Phi_{0}} \phi \right) = {\text{div}} \left( |\psi|^{2} ( \nabla \chi + \dfrac{2 \pi}{\Phi_{0}} \mathbf{A} ) \right).
	\end{equation}
	Using continuity equation \eqref{eq:nabla:j} together with \eqref{eq:l_E}, Eq.~\eqref{eq:app:A:TDGL:psi:Im} can be written as:
	\begin{equation}\label{eq:app:A:mu:nabla:E}
		\dfrac{|\psi|^{2}}{|\psi_{\infty}|^{2}} \mu = l_{E}^{2} \dfrac{2 \pi c}{\Phi_{0}} \nabla \mathbf{E},
	\end{equation}
	Further, dividing Eq.~\eqref{eq:app:A:TDGL:psi:Im} by $|\psi|^{2}$ and taking the time derivative, we obtain:
	{
		\begin{equation}\label{eq:app:A:dt:mu}
			\begin{aligned}
				\dfrac{\partial \mu}{\partial t} = &D \left( \nabla^{2} \mu - \dfrac{2 \pi c}{\Phi_{0}} \text{div\,} \mathbf{E} \right) + \\ 
				+&\dfrac{D}{|\psi|^{2}} \left( \nabla |\psi|^{2} , \nabla \mu - \dfrac{2 \pi c}{\Phi_{0}} \mathbf{E} \right) + \\
				+&\dfrac{D}{ |\psi|^{2} } \left( \nabla \dfrac{\partial |\psi|^{2} }{\partial t} ,   \nabla \chi + \dfrac{2 \pi}{\Phi_{0}} \mathbf{A} \right).
			\end{aligned}
	\end{equation}}
	Assuming  $\xi^{2} \gg$ $l_{E}^{2}$ we neglect the terms containing $\nabla |\psi|^{2}$. 
	
	Equation \eqref{eq:app:A:dt:mu} is then simplified to a diffusion-type equation for $\mu$ with a source ${\text{div}\,} \mathbf{E}$:
	\begin{equation}\label{eq:app:A:diff:type:mu}
		\dfrac{\partial \mu}{\partial t} = D \nabla^{2} \mu - D \dfrac{2 \pi c}{\Phi_{0}} \nabla \mathbf{E}.
	\end{equation}
	Combining Eq.~\eqref{eq:app:A:mu:nabla:E} and $l_{E}^{2} \eqref{eq:app:A:diff:type:mu} / D$, we arrive at a closed-form equation for $\mu$:
	\begin{equation}\label{eq:app:Amu:in:time:domain}
		\tau_{E} \dfrac{\partial \mu}{\partial t} = l_{E}^{2} \nabla^{2} \mu - f^{2} \mu,
	\end{equation}
	where $f = |\psi|/|\psi_{\infty}| \approx f_{0} + f_{1}$ and $\tau_{E} = l_{E}^{2} / D$. From Eq.~\eqref{eq:mu} we find:
	\begin{equation}
		\mu (\rho \rightarrow 0) \rightarrow (\mathbf{v}, \mathbf{e}_{\varphi}) \dfrac{1}{\rho} \ .
	\end{equation}
	Replacing the time derivation $\partial_{t}$ to $\partial_{t} - (\mathbf{v}, \nabla)$ (see Eq.~\eqref{eq:moving:frame}) and using the Fourier transform \eqref{eq:Fourier}, Eq.~\eqref{eq:app:Amu:in:time:domain} can be written as:
	\begin{equation}\label{eq:app:A:mu:closed:form}
		l_{E}^{2} \nabla^{2} \tilde{\mu} = ( f_{0}^{2} - i \omega \tau_{E} ) \tilde{\mu},
	\end{equation}
	because $(\mathbf{v}, \nabla) \mu$ and $f_{1} \mu$ are the values of second order in the vortex velocity $\mathbf{v}$.
	
	In the integral $\eqref{eq:int:mu:t}$, only the $e^{\pm i \varphi}$ harmonics of $\tilde{\mu}$ remain, so we write:
	\begin{equation}
		\tilde{\mu} (\omega, \rho, \varphi) = \tilde{\mu}_{+1} (\omega, \rho) e^{i \varphi} + \tilde{\mu}_{-1} (\omega, \rho) e^{-i \varphi}
	\end{equation}
	and, using Eq.~\eqref{eq:app:A:mu:closed:form}, arrive the equation for $\tilde{\mu}_{\pm 1}$:
	\begin{equation}\label{eq:app:A:mu:pm:1}
		l_{E}^{2} (\nabla^{2} - \dfrac{1}{\rho^{2}}) \tilde{\mu}_{\pm 1} = (f_{0}^{2} - i \omega \tau_{E}) \tilde{\mu}_{\pm 1}
	\end{equation}
	with the boundary condition:
	\begin{equation}
		\tilde{\mu}_{\pm 1} (\rho \rightarrow 0) \rightarrow \pm \dfrac{i}{2} (\tilde{\mathbf{v}}, \mathbf{e}_{x} \mp i \mathbf{e}_{y}) \dfrac{1}{\rho},
	\end{equation}
	
	We solve equation \eqref{eq:app:A:mu:pm:1} and evaluate integral \eqref{eq:int:mu:t}, using the Bardeen--Stephen approximation: $\chi_{1} = 0$ for $\rho < \xi$ and $f_{0} = h(\rho / \xi {-1})$, where $h(\rho/\xi{-1})$ is the Heaviside step function \cite{TinkhamM2004}. For $\rho < \xi$ equation \eqref{eq:app:A:mu:nabla:E} reduces to $\nabla^{2} \phi = 0$ and has the solution $\tilde{\phi}_{\pm 1}(\omega,\rho) = - (\Phi_{0} / 2 \pi c) \rho C_{\text{in}, \pm 1} (\omega) $, where $C_{\text{in}, \pm 1}$ are arbitrary constants. Thus, for $\rho < \xi$, the function $\tilde{\mu}_{\pm 1}$ takes the form:
	\begin{equation}
		\tilde{\mu}_{\pm 1} =  \pm \dfrac{i}{2} (\tilde{\mathbf{v}}, \mathbf{e}_{x} \mp i \mathbf{e}_{y}) \dfrac{1}{\rho} + C_{\text{in}, \pm 1} \rho.
	\end{equation}
	The solution of equation \eqref{eq:app:A:mu:pm:1} for $\rho > \xi$ is
	\begin{equation}
		\tilde{\mu}_{\pm 1} = C_{\text{out},\pm1} K_{1} ( \dfrac{\rho}{l_{E}} \gamma),
	\end{equation}
	where $K_{1}(z)$ is the modified Bessel function of second kind, $\gamma = (1 + \omega^{2} \tau_{E}^{2} )^{1/4} e^{-i \arctan (\omega \tau_{E}) / 2}$, because $|\mathrm{Arg}(z)|<\pi/2$. Imposing continuity of the $\tilde{\mu}_{\pm 1}$ and their derivatives $\partial_{\rho} \tilde{\mu}_{\pm 1}$ at $\rho=\xi$, we determine the constants $C_{\mathrm{in/out},\pm 1}$ $(K'_{1} (z) = - K_{0} (z) - z^{-1} K_{1} (z))$:
	\begin{multline}
		\begin{pmatrix}
			C_{\text{in},\pm 1} \\ C_{\text{out}, \pm 1}
		\end{pmatrix} = \pm
		\begin{pmatrix}
			- K_{0} (\dfrac{\xi}{l_{E}} \gamma) \\ 2
		\end{pmatrix} \dfrac{i}{2} \dfrac{1}{\xi^{2}} (\tilde{\mathbf{v}} {\,\cdot\,} (\mathbf{e}_{x} \mp i \mathbf{e}_{y})) \times \\
		\times \dfrac{1}{ \dfrac{ \gamma }{l_{E}}  K_{0} (\dfrac{\xi}{l_{E}} \gamma ) + \dfrac{2}{\xi} K_{1} (\dfrac{\xi}{l_{E}} \gamma ) }
	\end{multline}
	and evaluate integral \eqref{eq:int:mu:t} $(K'_{0} (z) = - K_{1} (z))$:
	\begin{multline}\label{eq:app:A:int:mu:frac}
		\dfrac{1}{\pi} \int d^{2} \mathbf{r} (\mathbf{d}, \mathbf{e}_{\varphi}) \dfrac{f_{0}^{2}}{\rho} \tilde{\mu} =  ( \mathbf{d}, \tilde{\mathbf{v}} )  \dfrac{2 l_{E}^{2}}{\xi^{2} \gamma^{2}} \times \\
		\times \dfrac{ K_{0} (\dfrac{\xi}{l_{E}} \gamma) }{ K_{0} (\dfrac{\xi}{l_{E}} \gamma) + \dfrac{2 l_{E}}{\xi \gamma} K_{1} ( \dfrac{\xi}{l_{E}} \gamma ) }.
	\end{multline}
	Under the condition $\xi \gg l_{E}$, we can use the asymptotic form of Bessel functions $K_{0} (z) \approx K_{1} (z) \approx (\pi / 2 z)^{1/2} e^{-z}$, so that the fraction in the second line of Eq.~\eqref{eq:app:A:int:mu:frac} approximately equals the unity, and we obtain Eq.~\eqref{eq:int:mu:t}:
	\begin{equation}
		\dfrac{1}{\pi} \int d^{2} \mathbf{r} (\mathbf{d}, \mathbf{e}_{\varphi}) \dfrac{f_{0}^{2}}{\rho} \tilde{\mu} = (\mathbf{d}, \tilde{\mathbf{v}}) \dfrac{ \alpha_{\text{Ohm}} }{1 - i \omega \tau_{E}}.
	\end{equation}
	
	\section{Expression for the electric field induced by the vortex motion.}\label{Ap:B}
	
	We consider the expression for the supercurrent $\mathbf{j}_{s}$ \eqref{eq:TDGL:j}:
	\begin{equation}\label{eq:App:C:js}
		\mathbf{j}_{s} = - \dfrac{c}{4 \pi \lambda^{2}} \dfrac{|\psi|^{2}}{|\psi_{\infty}|^{2}} \left( \mathbf{A} + \dfrac{\Phi_{0}}{2 \pi} \nabla \chi \right).
	\end{equation}
	Switching to the moving frame \eqref{eq:moving:frame} and taking the time derivative to \eqref{eq:App:C:js}, we obtain the following equation in the London limit \cite{TinkhamM2004,KopninNB2001}:
	\begin{equation} 
		\dfrac{\partial \mathbf{j}_{s}}{\partial t} = - \dfrac{c}{4 \pi \lambda^{2}} \left( \dfrac{\partial \mathbf{A}}{\partial t} - \dfrac{\Phi_{0}}{2 \pi} (\mathbf{v}, \nabla) \nabla \chi \right),
	\end{equation}
	Expressing $\partial_{t} \mathbf{j}_{s}$ in terms of the electric field $\mathbf{E}$, we write:
	\begin{equation}\label{eq:App:C:js:with:rot:chi}
		\dfrac{\partial \mathbf{j}_{s}}{\partial t} = \dfrac{c^{2}}{4 \pi \lambda^{2}} \left( \mathbf{E} - \dfrac{\Phi_{0}}{2 \pi c} \nabla \mu - \dfrac{\Phi_{0}}{2 \pi c} [\mathbf{v} , [\nabla, \nabla \chi]] \right),
	\end{equation}
	where the term $[\nabla, \nabla \chi] = -2 \pi \mathbf{e}_{z} \sum_{n} \delta^{(2)} (\mathbf{r} - \mathbf{R}_{n})$, $\mathbf{R}_{n}$ is the n-th vortex position. Thus, the average value of $[\mathbf{v} , [\nabla, \nabla \chi]]$ is:
	\begin{equation}
		\dfrac{1}{S_{\text{film}}} \int\limits_{\text{film area}} [\mathbf{v} , [\nabla, \nabla \chi]] d^{2} \mathbf{r} = - 2 \pi N [\mathbf{v}, \mathbf{e}_{z}],
	\end{equation}
	where $N$ is the total number of vortices and $S_{\text{film}}$ is the film area.
	
	Assuming that the superconductor contains vortices arranged in the Abrikosov lattice whose characteristic cell size exceeds $\xi$, we integrate \eqref{eq:App:C:js:with:rot:chi} over the sample, decomposing the integral into an average over a single unit cell of the Abrikosov lattice and a sum over all cells (term $\nabla \mu$ vanishes upon averaging):
	\begin{equation}
		\dfrac{\partial \langle \hspace{0.27ex} \mathbf{j}_{s} \rangle}{\partial t} = \dfrac{c^{2}}{4 \pi \lambda^{2}} \left( \langle \mathbf{E} \rangle + \dfrac{1}{c} [\mathbf{v}, \langle \mathbf{B}_{v} \rangle] \right),
	\end{equation}
	where $ \langle \mathbf{B}_{v} \rangle = ( \Phi_{0} N / S_{\text{film}} ) \mathbf{e}_{z}$. Upon Fourier transforming Eq.~\eqref{eq:Fourier}, we obtain the expression
	\begin{equation}
		- i \omega \langle \hspace{0.27ex} \tilde{\mathbf{j}}_{s} \rangle = \dfrac{c^{2}}{4 \pi \lambda^{2}} \left( \langle \tilde{\mathbf{E}} \rangle + \dfrac{1}{c} [\tilde{\mathbf{v}}, \langle \mathbf{B}_{v} \rangle] \right),
	\end{equation}
	which {coincides} with Eq.~\eqref{eq:induced:E}.
	
	\bibliography{refs-final-fixed}
	
\end{document}